\documentclass[pdflatex,sn-mathphys-ay]{sn-jnl}% Math and Physical Sciences Author Year Reference Style
\usepackage{graphicx}%
\usepackage{multirow}%
\usepackage{amsmath,amssymb,amsfonts}%
\usepackage{amsthm}%
\usepackage{mathrsfs}%
\usepackage[title]{appendix}%
\usepackage{xcolor}%
\usepackage{textcomp}%
\usepackage{manyfoot}%
\usepackage{booktabs}%
\usepackage{algorithm}%
\usepackage{algorithmicx}%
\usepackage{algpseudocode}%
\usepackage{listings}%
\theoremstyle{thmstyleone}%
\theoremstyle{thmstyletwo}%

\theoremstyle{thmstylethree}%

\begin{document}

\title[Multi-criteria ranking method]{Ranking honeybee breeder colonies: a probabilistic method using vine copulas to consider multiple criteria}


\author*[1]{\fnm{Christophe} \sur{Reype}}\email{chris.reype@gmail.com}
\author[2]{\fnm{Thomas} \sur{ Opitz}}
\author[3]{\fnm{Eleanora} \sur{ Bassi}}
\author[1]{\fnm{Benjamin} \sur{ Basso}}

\author[4,5]{\fnm{Charlène} \sur{ Dumas}}

\author[3]{\fnm{Massimiliano} \sur{ Gotti}}
\author[3]{\fnm{Giovanni} \sur{ Guido}}

\author[4,5]{\fnm{Guillaume} \sur{ Kairo}}

\author[8]{\fnm{Coline} \sur{ Kouchner}}

\author[4,5]{\fnm{Pascal} \sur{ Jourdan}}
\author[4,5]{\fnm{Pierre-Alban} \sur{ Mochet}}
\author[3]{\fnm{Michele} \sur{ Tagliabue}}
\author[1,4]{\fnm{Fanny} \sur{ Mondet}}


\affil*[1]{\orgdiv{UR406 Abeilles et Environnement}, \orgname{INRAE}, \orgaddress{ \city{Avignon}, \postcode{84914},\country{France}}}
\affil[2]{\orgdiv{UR546 BioSP}, \orgname{INRAE}, \orgaddress{ \city{Avignon}, \postcode{84914},\country{France}}}
\affil[3]{\orgname{UNAAPI}, \orgaddress{\city{Florence}, \country{Italy}}}
\affil[4]{\orgname{UMT ETTAP}, \orgaddress{\city{Avignon}, \postcode{84914},\country{France}}}
\affil[5]{\orgname{ADAPI}, \orgaddress{\city{Aix-en-Provence}, \country{France}}}

%%==================================%%
%% Abstract %%
%%==================================%%

\abstract{
Ranking honeybee colonies according to multiple performance criteria is a critical first step to selection in bee breeding, but deterministic scores provide limited information about how a selected colony compares with the population of available colonies. We propose a population-calibrated probabilistic score based on a weighted Euclidean distance to an ideal colony and an estimation of the joint distribution of the data. For a candidate colony, the probability-score is the probability that a colony drawn from a reference population outperforms the candidate. We estimate the dependence structure among $4$ chosen criteria using a vine copula after apiary-specific rank transformations, allowing different marginal distributions while modelling a common dependence structure. Importantly, the joint distribution calibrates the distance into a probability but does not change the ordering induced by a fixed distance and weight vector. We apply the method to $265$  colony observations from eight French and Italian apiaries during the 2024 and 2025 seasons. Dependence among the four criteria is weak in this dataset, and the vine-copula and independence probability-scores are therefore similar. A simulation study evaluates probability-score estimation under a range of dependence structures, while bootstrap and apiary-subset analyses quantify sensitivity to sampling and study composition. We additionally assess preference robustness by evaluating the proportion of admissible weight vectors for which a colony belongs to the top ranking. The framework provides an interpretable and reproducible population-relative score for multi-criteria colony selection.}




\maketitle

\section{Introduction}

The changes in climatic conditions induce an increase in the cost for the beekeepers, due to a reduction in productivity for honeybees such as \emph{Apis mellifera}~\citep{TosLom26} or the spread of invasive parasite species such as \emph{Varroa destructor}~\citep{NoeLec20}. The objective of selection is to enhance colony performance based on several traits or criteria concomitantly such as honey production, colony strength, hygienic behaviour, or resistance to \emph{Varroa destructor}. Selection on several traits leads to a multi-criteria decision problem, where several variables of different nature and scale must be considered jointly. This work aims to  present a method usable by a beekeeper, to rank the colonies of their apiary based on their performance in the field.

In the literature, the ranking of individuals is made using multi-criteria decision-making (MCDM) methods, including weighted sum models \citep{MarAro09}, the Technique for Order Preference by Similarity to Ideal Solution (TOPSIS) \citep{HwaYoo81,CelTuy20}, and Pareto-front approaches \citep{GaniAsj21}. Weighted sum models are relatively easy to implement and rely on the relative importance of the considered variables given by the user to define their weights. The TOPSIS method compares each individual to the perfect individual, a theoretical individual with the best values for each of the considered variables, and to the worst individual, a theoretical individual with the worst values for each of those variables. Pareto-based approaches identify the set of non-dominated individuals, \emph{i.e.}, those for which no other individual is better on all criteria. While these methods provide deterministic rankings, they generally lack an interpretation of the comparison between individuals and do not investigate the joint distribution of the data. Because the considered variables may be dependent, modelling their joint behaviour can improve the population calibration and interpretation of a multi-criteria score, even when the underlying distance-based ordering remains unchanged.

Specifics methods are developed for ranking and selecting individuals for animal breeding. These methods associate a score to each individual according to one or multiple criteria. The main method used in selection is the Best Linear Unbiased Predictor (BLUP)~\citep{Che77}. For each individual (\emph{i.e.} each colony), a breeding value is estimated. This value takes into account \emph{a posteriori} knowledge such as the performance of the individual in one or multiple criteria, and \emph{a priori} knowledge such as information on the pedigree and genetic relatedness between individuals. In apiculture, the individual is the combination of the queen and the workers, thus the performance is assimilated as a phenotypical value of the colony~\citep{BieEhr07}. 
A selection based only on the performance of the individuals is a massal selection. Such selection is made using either regression analyses or a score~\citep{BucAnd13}. Thus, selection relies either on the performance of the pedigree or on a phenotypic value based score not easily interpretable.

The aim of this paper is to propose a tool for massal selection directly usable and interpretable by beekeepers. The individuals (\emph{i.e.} colonies) are ranked according to a probability-based score, that combines MCDM concepts with a statistical model of the joint distribution of the data. For each colony, the weighted Euclidean distance to a perfect colony, a theoretical colony with the best values for the considered variables, is computed. The joint distribution of the data is modelled with a vine copula, which captures complex pairwise dependencies while remaining tractable in multi-dimensions. For a given real colony, the probability-score is defined as the probability that a randomly generated individual is closer to the perfect colony. A better real colony has a lower probability-score thus a lower probability to find a better colony. The probability-score can also be interpreted as the probability that the colony is not the best colony possible. The difference between probability-scores  provides an intuitive measure of comparison between colonies: choosing one colony over another will increase or reduce the probability of selecting the best possible colony, depending on the sign of the difference.

The method is built on a dataset comprising $265$ colonies observed one year spread over eight apiaries (four in France, four in Italy) and two seasons. Three weighting strategies are examined: \textbf{uniform}, \textbf{productivity-focused}, and \textbf{\emph{Varroa}-focused}. The latter two respectively prioritize high honey yield or traits correlated to \emph{Varroa} resistance, while the uniform strategy favours generalist colonies. The approach is validated and shown to be robust via envelope tests and bootstrap analyses. %The remainder of the paper is organised as follows. Section~\ref{sec:matAndMet} describes the data, the vine-copula model, and the risk-based ranking procedure. Section~\ref{sec:results} presents the empirical results and robustness checks. Section~\ref{sec:discussion} discusses the implications and limitations, and Section~\ref{sec:conclusion} concludes.

\section{Materials and Methods}

\label{sec:matAndMet}

\subsection{Data presentation}

\label{subsec:data_presentation}

The study draws on data collected during the Melior'Api project, which monitored honey-bee colonies over two consecutive beekeeping-seasons (2024/2025 and 2025/2026) in eight apiaries, four in the Provence region of South-eastern France and four in the Piedmont region of North-western Italy. Each apiary began the season with $60$ colonies, yielding a total of $960$ colonies monitored ($480$ per season). Unique identifiers of the apiaries are constructed using "FR$\_$" and "IT$\_$", respectively for France and Italy, and two initial letters to identify the beekeeper and the year of the monitored season.

A beekeeping season is divided into five monitoring events called visits as described in Table~\ref{tab:visites}. The first monitoring event, or visit, $V1$ happens at the beginning of spring. The beekeeper chooses the colonies to monitor during the season. The visit $V2$ takes place before the honey production of reference (lavender in France and chestnut in Italy). The visit $V3$ is scheduled after the honey production of reference and allows the estimation of the production of each colony. The visit $V4$ occurs at the beginning of autumn. During this visit, the colonies are treated against \emph{Varroa} and the food autonomy of the colonies are determined. The visit $V5$ is after the end of winter and allows determining the winter survival and the remaining colonies to reproduce. If an apiary is monitored the following beekeeping-season, the monitoring events start again at $V1$. In this case, monitored event are distinguished by the year of the season.

\begin{table}[t]
\centering
\caption{Description of the five monitoring events.}
\begin{tabular}{p{0.07\textwidth}*{2}{p{0.41\textwidth}}}
  \hline
Name & Time & Description\\
  \hline
$V1$ & Beginning of spring & Period to determine the colonies to monitor\\

$V2$ & Before the honey production of reference & \\

$V3$ & After the honey production of reference & Period to determine the productivity of the colonies\\

$V4$ & Beginning of autumn  & Period to treat the colony and estimate the food autonomy\\

$V5$ & End of winter & Period to determine the winter survival and reproduce the chosen colonies\\
\hline
\end{tabular}

\label{tab:visites}
\end{table}

At each visit, several variables are systematically measured, such as the amount of bees, the amount of brood, and the amount of \emph{Varroa destructor} (see the dataset available at \url{https://forge.inrae.fr/christophe.reype/Inrae_Bees}), but missing data can arise. Because this work relies on the estimation of a multivariate probability density by a copula, we require a minimum of $50$ different individuals in the considered dataset per variable to estimate parameters. Expert judgement and considerations of data availability led to the choice of ranking the colonies according to four criteria measured by $K=4$ variables (see Table~\ref{tab:criteria}): honey yield, colony strength (bee population), \emph{Varroa} infestation level, and hygienic behaviour (pin-test). The pin-test is made by puncturing $50$ broods cells with a pin. The pin-test score is then measured by the ratio of closed broods cells among these $50$ after a period of  three and a half hours.

The honey yield in kg, noted $Harvest$, is measured by the difference between the weight of the colony at the end of the honey production of reference and its weight after the harvest. The colony strength is measured by the number of bees after the winter ($V5$) evaluated with the ColEval method~\citep{HerMai20} and noted $Strength$. 
%A colony with a high value of $Strength$ will start the season strongly. 
The \emph{Varroa} infestation is measured by $Infestation$, the difference in the proportion of \emph{Varroa} mites in a sample of bees collected before the honey production of reference ($V2$) and at the beginning of autumn ($V4$) measured as in~\citep{DieNaz12}.
 %A colony with a high value of $Infestation$ is able to slow the spreading of \emph{Varroa}.
  The hygienic behaviour is measured by the maximum pin-test score measured as in~\citep{DieNaz12} during the season and is noted $Hygiene$. In the season, two pin-tests are measured, one at $V2$ and the other at $V4$. The hygienic behaviour is linked to the resistance against \emph{Varroa}. For these four criteria, a high value indicates a colony with a good performance. Thus, a colony whose performance is considered as good  will ideally have a relatively high value for each of these four criteria: a high amount of honey produced, a high number of bees after winter, a low increase of \emph{Varroa} mites between visits $V2$ and $V4$, and a high maximum pin-test score.

\begin{table}[t]
\centering
\caption{Description of the $K=4$ criteria.}
\begin{tabular}{p{0.12\textwidth}p{0.6\textwidth}p{0.12\textwidth}p{0.12\textwidth}}
  \hline
Name  & Description & Unit & Support\\
  \hline
$Harvest$ & Total honey produced & kg & $\mathbb{R}_+$\\

$Strength$ & Number of bees surviving winter ($V5$) evaluated with the ColEval method~\citep{HerMai20} & count & $\mathbb{N}$\\

$Infestation$ & The difference in number of \emph{Varroa} mites before the honey production of reference ($V2$) and at the beginning of autumn ($V4$) measured as in~\citep{DieNaz12} & proportion & $\mathbb{R}$\\

$Hygiene$ & Maximum pin-test score (hygienic behaviour) measured during the season~\citep{NewOst86} & score & $[0,100]$\\
\hline
\end{tabular}

\label{tab:criteria}
\end{table}

Table~\ref{tab:effectifExit} gives the total number of colonies at each visit. There are three reasons for the colonies to exit the experiment: death, change of queen, or early treatment against \emph{Varroa}. A change of queen results in a change of the bees' genetics. Because there is no evidence that the phenotype of the colony is the same throughout the season, such colonies are withdrawn from the experiment. One of the objectives of the Melior'Api project is to improve the resistance against \emph{Varroa}. An early treatment against \emph{Varroa} (before $V4$) indicates that the colony was not able to slow the spreading of the parasite. Moreover, a colony with an early treatment does not compete fairly against a colony without it. Thus, such colonies are withdrawn from the experiment.

\begin{table}[t]
\centering
\caption{Total number of colonies at each visit.}
\begin{tabular}{c*{9}{c}}
  \hline
Visit & 2024 & 2025 & Total \\
\hline
$V1$   &     468            &             480             &         948\\
$V2$   &      402           &              409            &          811\\
$V3$   &     336           &              341             &         677\\
$V4$   &     209           &              309            &          518\\
$V5$   &     136           &              272             &         408\\

   \hline
\end{tabular}

\label{tab:effectifExit}
\end{table}

After removing the colonies with missing values, the number of colonies retained is $m=265$ and the number of colonies retained for each apiary, noted $m_\ell$ for apiary $\ell$, is given in Table~\ref{tab:effectif}. During the season $2024$, the number of bees at the end of winter were not measured in Italy. Moreover, during the season $2025$, an italian beekeeper and a french beekeeper choose to withdraw their apiary from the experiment due to a high mortality in their colonies.
% one apiary in Italy and one in France have exited the experiment due to a lack of follow-up from the beekeepers. %In this work, the choice is made to keep the predefined criteria and all apiaries to study the method in real conditions.

\begin{table}[t]
\centering
\caption{Number of colonies retained in each apiary ($m_{\ell}$). The total number of colonies used in this paper is $m=265$.}

\begin{tabular}{cc||cc}
  \hline
Apiary& $m_\ell$&Apiary& $m_\ell$\\
  \hline
FR$\_$BC$\_$2024 &  41& FR$\_$PM$\_$2024 &  30\\  
FR$\_$FV$\_$2024 &  8 &FR$\_$PM$\_$2025 &47  \\ 
FR$\_$FV$\_$2025 &  23 &IT$\_$CT$\_$2025 & 22\\  
FR$\_$OA$\_$2024 &  17 &IT$\_$EL$\_$2025 &21\\ 
FR$\_$OA$\_$2025 &   39 &IT$\_$YC$\_$2025 & 17\\

\hline

\end{tabular}

\label{tab:effectif}
\end{table}

The total number of colonies $m$ leads to two considerations. The first consideration is on setting a minimal number of colonies per apiary. Specifically, the interest of apiary FR$\_$FV$\_$2024 and its $8$ colonies has been questioned. The second consideration is on the choice of the criteria. Changing the criteria can increase the total number of colonies used, $m$, up to $408$. Even if an increase in the number of individuals should improve the estimation and the results of the statistical tools, we decided to keep the current criteria and all apiaries. This choice was made to confront the method to real conditions such as the application of the method on a small apiary or with a dataset that may not be large enough to allow for a robust estimation of the distribution of the data.

The project Melior'Api considers three trade-offs of the chosen criteria, called strategies (see Table~\ref{tab:strategies}). The first strategy, called \textbf{uniform} strategy, has no preference for either criterion. This strategy favours generalist colonies: colonies with a good compromise of values for all criteria. The second strategy is heavily biased toward productivity. The best colonies are specialists at producing honey. The third strategy puts emphasis on the capacity to resist \emph{Varroa}. In practice, each strategy is represented as a convex sum: a weight $w_i$ between $0$ and $1$ is associated to each criterion and the weights sum to $1$. A weight can be interpreted as the importance of its criterion. For example, in the productivity-focused strategy, $Harvest$ has an importance of $70\%$ and the other criteria $10\%$ each. 

\begin{table}[t]
\centering
\label{tab:strategy}
\caption{Description of the chosen ranking strategies.}
\begin{tabular}{l*{4}{c}}
  \hline
Strategy & Harvest &Strength & Infestation &Hygiene\\
  \hline
1 - \textbf{Uniform}&0.25&0.25&0.25&0.25\\

2 - \textbf{Productivity-focused}& 0.7&0.1&0.1&0.1\\

3 - \textbf{\emph{Varroa}-focused}&0.15&0.15&0.35&0.35\\
   \hline
\end{tabular}

\label{tab:strategies}
\end{table}

\subsection{Notation and marginal standardisation}

\label{subsubsec:notation}

The colonies are represented as points in a $K$-dimensional space (with $K=4$ in our study), called the data space and denoted $\Omega\subseteq\mathbb{R}^{K}$. The coordinates of colony $1\leq i\leq m$ are the measurements of the $K$ variables and are written $\mathbf{x}^{(i)}=(x^{(i)}_1,\dots,x^{(i)}_K)\in\Omega$. 

Because the measurement units and scales of the four variables differ, we propose two standardisations to make them commensurable, \emph{i.e.} to cast them into a common scale. The first standardisation, called \textbf{min-max scaling}, scales the observed range to the unit interval $[0,1]$: 
\begin{equation}
u^{(i)}_k=\frac{x^{(i)}_k-min_j\{x^{(j)}_k\}}{max_j\{x^{(j)}_k\}-min_j\{x^{(j)}_k\}}\in [0,1].
\label{eq:minMaxScaling}
\end{equation}
This standardisation is not robust to extreme values, but the relative spacings between values are conserved.

A second standardisation, called \textbf{rank scaling}, derived from the copula theory is also considered and allows for probabilistic interpretation as the value of the empirical cumulative distribution function:
\begin{equation}
u^{(i)}_k=\frac{\mathrm{pos}(x^{(i)}_k)}{m_\ell+1}\in [0,1],
\label{eq:rankScaling}
\end{equation}
where $\mathrm{pos}(x^{(i)}_k)$ is the position of the colony $i$ in the apiary $\ell$, in ascending order. This standardisation is robust to extreme values and outliers. Here, the relative spacing between original values is lost.

\subsection{Copula models}

\label{sec:copulaModels}

Modelling biological data often requires estimating multivariate distributions for which linear correlation fails to capture complex dependencies between variables. Copulas address this challenge by separating the marginal behaviour from the dependence structure: each marginal distribution is modelled independently, and a copula couples them to form the full multivariate distribution. To allow for high flexibility, the copula can be constructed as a cascade of bivariate copulas.

Let $X=(X_{1},\dots,X_{K})^{\!\top}$ be a $K$-dimensional continuous random vector with joint cumulative distribution function (cdf) $F$, joint density $f$, and marginal cdfs $F_{k}$ with corresponding marginal densities $f_{k}$ for $k=1,\dots,K$.

By the fundamental Sklar's theorem~\citep{Sklar59}, there exists a function $C\!:\![0,1]^{K}\rightarrow[0,1]$, called the copula, and its density $c$, such that
\begin{equation}
F(x_{1},\dots,x_{K}) = C\left(F_{1}(x_{1}),\dots,F_{K}(x_{K})\right),
\label{eq:copula}
\end{equation}
and
\begin{equation}
f(x_{1},\dots,x_{K})
=
c\left(F_{1}(x_{1}),\dots,F_{K}(x_{K})\right)\,
\prod_{k=1}^{K} f_{k}(x_{k}).
\label{eq:copulaDensity}
\end{equation}
The copula $C$ is unique when all marginal cdfs $F_{k}$ are continuous.

The representation above uses a single $K$-variate function to capture all dependence among the marginals. In higher dimensions where the dependence structure may be quite complex, a vine copula~\citep{BedCoo02} constructs a flexible multivariate copula recursively from a collection of bivariate copulas (pair-copulas). For a $K$-dimensional problem, up to $K(K\!-\!1)/2$ pair-copulas are combined, where each pair can have one or several parameters to estimate and models a specific aspect of the dependence, by contrast with various simpler standard copula models with only a single parameter and exchangeability of all $K$ variables.

For $K=4$, a common pair-copula decomposition (up to a permutation of the variables) can be written as
\begin{align}
f(x_{1},x_{2},x_{3},x_{4})
=&
c_{(1,4)|(2,3)}\left(F\left(x_{1}| x_{2},x_{3}\right),
 F\left(x_{4}| x_{2},x_{3}\right)\right)   \nonumber\\
&\times c_{(1,3)|2}\left(F(x_{1}| x_{2}),F(x_{3}| x_{2})\right)\,
   c_{(2,4)|3}\left(F(x_{2}| x_{3}),F(x_{4}| x_{3})\right) \nonumber\\
&\times c_{(1,2)}\left(F(x_{1}),F(x_{2})\right)\,
   c_{(2,3)}\left(F(x_{2}),F(x_{3})\right)\,
   c_{(3,4)}\left(F(x_{3}),F(x_{4})\right) \nonumber\\
&\times f_{1}(x_{1})\,f_{2}(x_{2})\,f_{3}(x_{3})\,f_{4}(x_{4}),
\label{eq:pairCopulaDecomposition4D}
\end{align}

Thus, six pair-copulas are used to capture the dependence structure.

By choosing a particular vine structure (among the $K!/2$ distinct $D$-vines~\citep{AasCzaFriBak09}) and selecting an appropriate parametric family for each pair-copula, one can approximate a wide variety of complex multivariate dependencies.

In the present study, vine structures are estimated in \texttt{R} 4.5.2 using the \texttt{rvinecopulib} package (version 0.7.3.1.0). The default sequential algorithm based on empirical Kendall's $\tau$~\citep{DissBreCzaKur13} is employed to identify the permutation of variables and the conditional dependence structures: at each tree level the conditioning set is enlarged, and the pair-copula that maximises the sum of absolute empirical Kendall's $\tau$ is selected, thereby capturing the strongest pairwise dependencies first. For each selected pair, the copula parameters for different possible copula-family specifications are estimated by maximum likelihood, and the copula family is chosen by Akaike Information Criterion (AIC).

\subsection{Construction of a ranking criterion}

\subsubsection{Sequential ranking criterion}

We compare our newly proposed method to a simple and commonly applied method of selection consisting of first keeping only the individuals that have the top values for any of the criteria. Thus, the selection is typically among specialist individuals: individuals that show very good performance for at least one criterion. Then, these individuals are ranked in the descending order of their worst criterion. 

In practice, few, if any, individuals simultaneously show top-ranking performance for all criteria. For example, if only the top $5$ individuals for at least one criterion among $K=4$ criteria are kept, this filtering yields a candidate set whose size lies between $5$ and $K\times 5 = 20$.

\subsubsection{Ranking criterion based on a weighted distance}

\label{subsubsec:weightD}

To formalise the ranking of colonies in our method, we introduce the \textbf{perfect colony}: a theoretical colony that possesses the best observed value for each criterion in the considered apiary, \emph{i.e.} the perfect colony is the component-wise maximum of the set of all considered colonies. It corresponds to the maximum values that are also explored in the sequential ranking presented above. The perfect colony may not exist in reality, but it provides a reference point for comparison. 

In the following, the data are supposed standardised by either the min-max scaling and the rank scaling. In both cases, the perfect colony is set as $\mathbf{u}^*=(1,\dots,1)$: in practice, the perfect colony is usually always better than the colony considered as being the best one of the apiary according to some criteria.% Among two colonies, the relatively better one must be "more similar" to the perfect colony, \emph{i.e.} have lower distance to the perfect colony. 

The Euclidean distance to the perfect colony grows quadratically as a standardised measurement deviates from $1$. Small deviations from $1$ increase the distance only marginally, whereas larger deviations can contribute relatively more strongly to the distance. By introducing weights to each criterion, that is to set the relative importance of each criterion, the weighted Euclidean distance allows for the calibration of the ranking according to the beekeeper's preferences. The weighted Euclidean distance of the colony numbered $i$ with the weights $w=(w_1,\cdots,w_K)$ is:
\begin{equation*}
D(\mathbf{u}^{(i)})=d(\mathbf{u}^{(i)},\mathbf{u}^*)=\sqrt{\sum_{k=1}^K w_k(u^{(i)}_k-1)^2},
\end{equation*}
If the weights satisfy $w_{k}\ge 0$ and $\sum_{k=1}^{K}w_{k}=1$, each weight $w_{k}$ can be interpreted as the proportion of importance attached to criteria $k$. This distance can be used to select either generalist or specialist colonies (see Table~\ref{tab:strategies}). Let $d_{(j,m)}$ be the ordinal statistic such as $d_{j,m}$ is the $j$-th smallest distance among the $m$ distance, then the first colony in the ranking is the one which $D(\mathbf{u}^{(i)})=d_{(1,m)}$ and the last colony in the ranking is the one which $D(\mathbf{u}^{(i)})=d_{(m,m)}$.

\begin{figure}[t]
\centering
\includegraphics[scale=0.2]{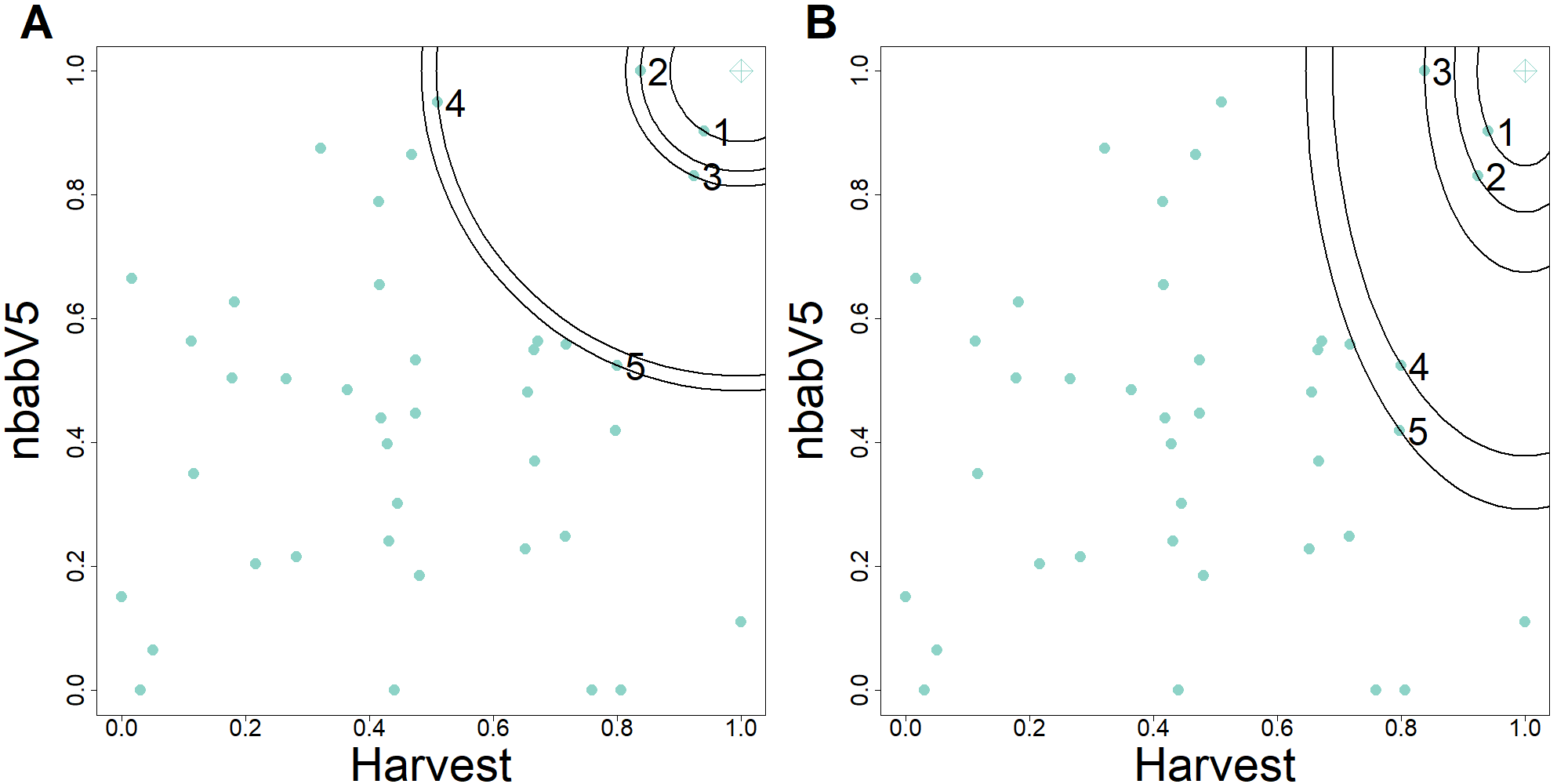}
\caption{Effect of the strategy ($w=(0.5,0.5)$ - A, $w=(0.8,0.2)$ - B) on the projection of the apiary FR$\_$BC$\_$2024 in the bi-dimensional space formed by $Harvest$ and $Strength$ with the rank scaling. The rank and the equidistant points of the top $5$ colonies are indicated.}
\label{fig:effectPond}
\end{figure}

Figure~\ref{fig:effectPond} is an example of two choices of weights. In this bivariate display only $Harvest$ and $Strength$ are considered. The colonies of the apiary FR$\_$BC$\_$2024 are ranked according to two strategies: a uniform strategy where each variable has the same importance $w=(w_1,w_2)=(0.5,0.5)$ (left column) and a strategy emphasising $Harvest$ $w=(0.8,0.2)$ (right column). In each case, the colonies are represented by dots and the perfect colony by a symbol set at coordinates $(1,1)$. For the first strategy, points equidistant to the perfect colony are on a circle centred on the perfect colony. For the second strategy, point equidistant to the perfect colony are on an ellipse centred on the perfect colony. Because $Harvest$ is more important, it should be high, whereas $Strength$ may be lower. The results with the min-max scaling are presented in the Appendix.

\subsubsection{Ranking criterion based on the probability of having a better colony}

\label{subsubsec:risk}

The aim of this section is to improve interpretability and reduce statistical uncertainty of the distance ranking method, so the ranking remains intuitive even for readers without a mathematical background.

In the standardised data space $[0,1]^{K}$, a point $\mathbf{t}$ is better than colony $\mathbf{u}^{(i)}$ if its weighted distance to the perfect colony does not exceed that of $\mathbf{u}^{(i)}$:
\begin{equation*}
D(t)=d(\mathbf{t},\mathbf{u}^*)\leq d(\mathbf{u}^{(i)},\mathbf{u}^*)=D(\mathbf{u}^{(i)}).
\end{equation*}
The set of points that are at least as good as colony $i$ is therefore  
\begin{equation*}
A(\mathbf{u}^{(i)},\mathbf{u}^*)=\{\mathbf{t}\in[0,1]^K,\,D(t)\leq D(\mathbf{u}^{(i)})\}.
\end{equation*}

Lets $f:[0,1]^K\longmapsto\mathbb{R}_+$ denote the unknown joint density of the standardised data. The theoretical probability that a randomly drawn colony is at least as good as colony $i$ is  
\begin{equation}
r(\mathbf{u}^{(i)})=\mathbb{P}\left(A(\mathbf{u}^{(i)},\mathbf{u}^*)\right) = \int_{\mathbf{t}\in[0,1]^K}  f(\mathbf{t})\mathbf{1}\left\{\mathbf{t}\in[0,1]^K,\,D(t)\leq D(\mathbf{u}^{(i)})\right\}\,d\mathbf{t}.
\label{eq:riskTheoretical}
\end{equation}
The quantity $r(\mathbf{u}^{(i)})$, called probability-score, can be interpreted as the probability that the colony can be outperformed. This interpretation is made under the assumption that $f$ describes the distribution of the colonies and the considered strategy is indeed the good strategy.

Smaller probability-score values indicate better colonies. The difference between two probability-scores quantifies the difference in the probability of being outperformed. This change should be left to the discretion of the beekeeper.

Lets the function $F_D:[0,1]\longmapsto[0,1]$ defined as
\begin{equation*}
\forall d\in\mathbb{R}:\, F_D(d)=\mathbb{P}\left(D(u)\leq d\right),
\end{equation*}
be the  probability that a random colony is closer than a distance $d$. Then the probability-score is
\begin{equation*}
r(\mathbf{u}^{(i)})=F_D(\mathbf{u}^{(i)}).
\end{equation*}
Thus, if $F_D$ is strictly increasing then
\begin{equation}
D(\mathbf{u}^{(i)}) \leq D(\mathbf{u}^{(j)}) \Leftrightarrow r(\mathbf{u}^{(i)}) \leq r(\mathbf{u}^{(j)}).
\end{equation}
The choice of $f$ changes the  value of the probability-score but not the ranking.

If there is no information on the distribution of the data, $f$ is estimated by a uniform distribution \emph{i.e.} an independence copula. Thus, the distance ranking criterion with the rank scaling can be transformed into a probability-score by, for the colony $i$:
\begin{equation*}
r(\mathbf{u}^{(i)})=\frac{1}{4}V(D(\mathbf{u}^{(i)})),
\end{equation*}
with $V(D(\mathbf{u}^{(i)}))$ the volume of the $K$-dimensional ellipsoid of radius $D(\mathbf{u}^{(i)})$. In the specific case of the uniform ranking strategy, the ellipsoid is a $K$-dimensional sphere, thus:
\begin{equation*}
r(\mathbf{u}^{(i)})=\frac{1}{4} \frac{\pi^{K/2}}{\Gamma(\frac{K}{2}+1)}(D(\mathbf{u}^{(i)}))^K,
\label{eq:riskIndependant}
\end{equation*}
with $\Gamma(\cdot)$ the Euler's gamma function.

For the selection, only individuals with high measurement in each criterion are interesting. These individuals are the results of extreme events. Thus, an empirical distribution or a uniform distribution is not enough to estimate the distribution of the data. We propose to estimate the distribution of the data by copulas.

In practice the true density $f$ is unknown and must be estimated. Let $\widehat{f}$ be the fitted density obtained from the vine-copula model and draw $n$ theoretical colonies $\mathbf{t}_1,\dots,\mathbf{t}_n$. The quantity defined in Equation~\eqref{eq:riskTheoretical} is then approximated by a Monte-Carlo estimator, which attains high accuracy for large $n$ with a low computational cost:
\begin{equation}
\widehat{r}(\mathbf{u}^{(i)})=\frac{1}{n}\sum_{j=1}^n \mathbf{1}\left\{d(\mathbf{t}_j,\mathbf{u}^*)\leq d(\mathbf{u}^{(i)},\mathbf{u}^*)\right\}.
\label{eq:riskEmpirical}
\end{equation}

In general, this probability-score measure implicitly assumes that the fitted density $\widehat{f}$ faithfully represents the true joint distribution of the traits, and that the weight vector $\mathbf{w}$ used in the distance $d(\cdot,\cdot)$ correctly reflects the beekeeper's true preferences. The probability-score calibrates the distance relative to the reference population and provides an interpretable probabilistic measure of how many reference-population colonies are expected to perform at least as well as the candidate. 

\subsubsection{Practical estimation of the probability-score}

\label{subsubsec:tutorial}

The joint distribution of the data is modelled with a vine copula (Section~\ref{sec:copulaModels}).  
Because vine-copula estimation requires a reasonably large sample to identify parameters, only considering one apiary with a maximum of $47$ hives in the dataset is not enough. Thus, we decide to pool all apiaries. 

Because apiaries may differ in size and marginal distributions, we first map each apiary's position to a common uniform copula-scale $[0,1]$ with an apiary-specific transformation. For the colony $i$ of a given apiary $\ell$, the corrected position is:
\begin{equation}
\mathrm{pos}_{C}(x^{(i)}_k)=\frac{\mathrm{pos}(x^{(i)}_k)-1}{m_\ell-1}m+1,
\label{eq:correctedPosition}
\end{equation}
For each apiary, the corrected position of the best colony is $m+1$ and $1$ for the worst colony.

To obtain comparable $\mathbf{u}$ vectors across apiaries, we then apply the rank scaling:
\begin{equation}
u^{(i)}_k=\frac{\mathrm{pos}_{C}(x^{(i)}_k)}{m+2} \,\in [0,1].
\label{eq:standardisationFinal}
\end{equation}
The shift by $2$ avoids exact ones, which can lead to boundary issues in the copula density.

The joint distribution of the standardised data is used to model the relative multivariate performance within apiaries not the absolute biological performance between apiaries. This is translated by the fact that a same colony may be good for an apiary but bad for another apiary.

The empirical probability-score $r(\mathbf{u}^{(i)})$ defined in Equation~\eqref{eq:riskEmpirical} is computed by Monte-Carlo simulation with $n=10^5$ theoretical colonies. 

We propose another score to distinguish between robust generalists and strategy-dependent specialists. Let $W$ be the set of all possible strategies. We define the score $P_{W,a}(i)$ as the ratio of strategies amongst $W$ for which the colony $i$ is in the top $a\in\mathbb{N}$ ranks. In practice, $W$ is approximated by the set of weights representing integer percentage, \emph{i.e.} by an approximate uniform distribution over the simplex (in $\mathbb{R}^K$) of possible strategies.  A low score indicates that the colony is a top ranker in just a few strategies, \emph{i.e.} a strategy-dependent specialist. A high score indicates that the colony is a top ranker in lot of strategies, \emph{i.e.} a robust generalist. For $K=4$, $W$ is numerically discretised by a set of $156903$ strategies.

If the real joint distribution and the resulting true probability-scores of the data are known, the quality of the estimated probability-score can be estimated with the root square mean error:
\begin{equation}
RMSE=\sqrt{\frac{1}{m}\sum_{i=1}^m \left(\widehat{r}(\mathbf{u}^{(i)})-r(\mathbf{u}^{(i)})\right)^2}
\label{eq:rmse}
\end{equation}
with $\widehat{r}(\mathbf{u}^{(i)})$ the probability-score computed on the estimated joint distribution and $r(\mathbf{u}^{(i)})$ the probability-score computed on the real joint distribution. A low root square mean error indicates a good estimation of the probability-score.

\section{Results}

\label{sec:results}

This section illustrates how the proposed methodology behaves on the data described in Section~\ref{subsec:data_presentation}.  

We first examine how the two types of marginal standardisation affect the probability-score-based ranking, and  then present the full vine-copula probability-score curves under the three weighting strategies, and finally assess robustness via bootstrap and envelope tests.

%section 1 : comparaison entre probability-score et distance
%section 2 : comment utiliser le probability-score en pratique
%section 3 : robustesse de la methode

\subsection{Data visualisation}

\begin{figure}[t]
\centering
\includegraphics[scale=0.4]{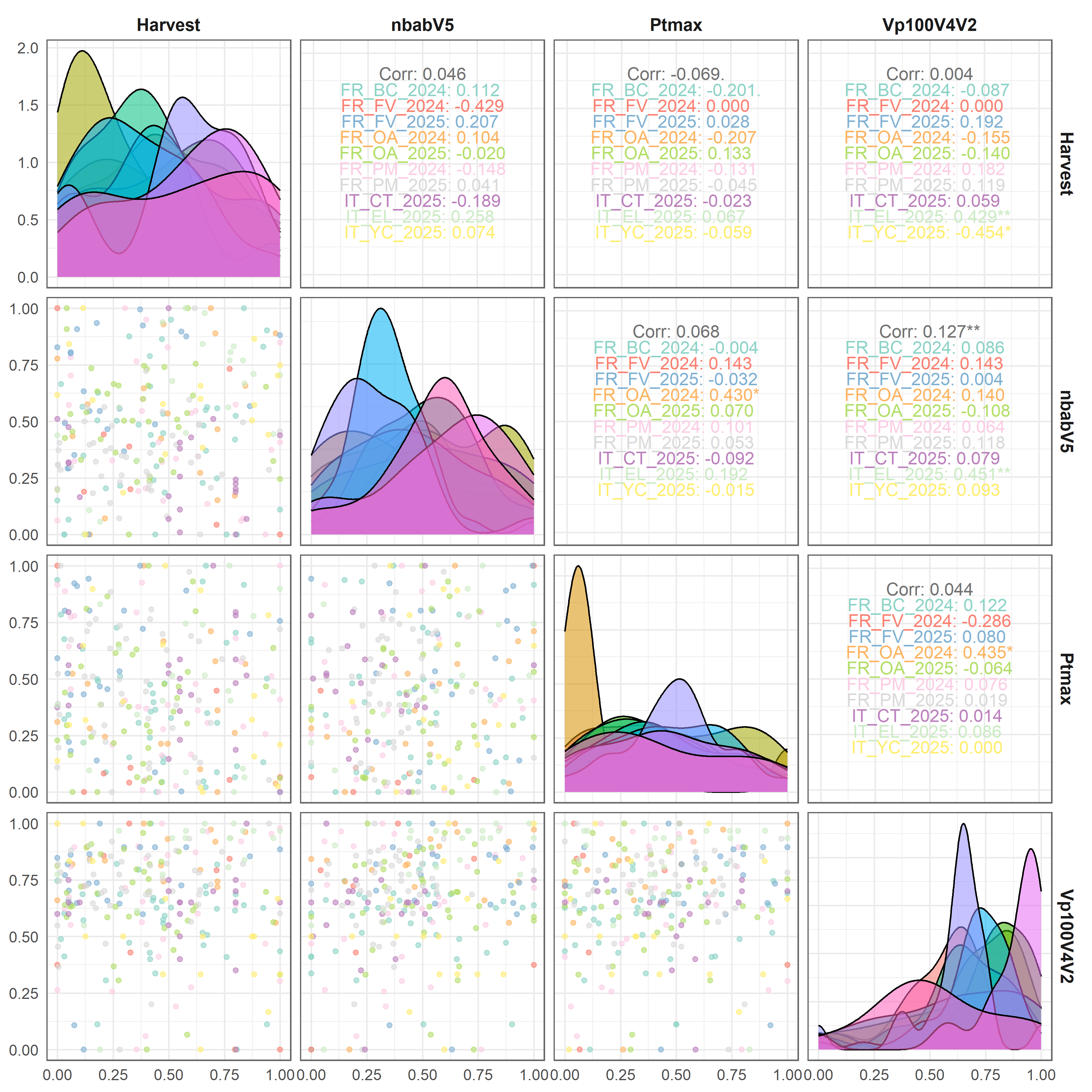}
\caption{Pair plot of the selected variables.  The lower-triangular panels display bivariate scatter plots. The diagonal panels show kernel-density estimates for each variable. In the upper diagonal, the Kendall's tau coefficient of the pooled apiaries is in black, each colour corresponds to an apiary.}
\label{fig:pairPlotCriteria}
\end{figure}

The pooled dataset is presented in the pair-plot of Figure~\ref{fig:pairPlotCriteria}. Plots on the diagonal are the kernel-smoothed marginal densities of each criterion per apiary. Lower-triangular panels are bi-dimensional projections of the data where each colour represents an apiary. Upper-triangular panels are Kendall's $\tau$ coefficient for the pooled dataset and per apiary. Only the correlation between the variation in the number of \emph{Varroa} mites ($Infestation$) and the number of bees after winter ($Strength$) is significant for the complete dataset. For some of the apiaries, the correlation between the variation in the number of \emph{Varroa} mites ($Infestation$) and the honey produced ($Harvest$), between the variation in the number of \emph{Varroa} mites ($Infestation$) and the pin-test score ($Hygiene$), and between the pin-test score ($Hygiene$) and the number of bees after winter ($Strength$) is also significant.

\subsection{Simulation study for the effect of the dependency structure}

The dependence structure in the data is relatively weak in the dataset (see~\ref{fig:pairPlotCriteria}). Before studying a fitted vine copula on the real data, we explore $8$ configurations for the dependency structures of synthetics data with $K=4$ variables:
\begin{itemize}
\item independence between Gaussian variables,
\item Gaussian dependence between Gaussian variables (weak - $\rho=0.2$, strong $\rho=0.8$),
\item Gumbel copula (weak - $\theta =1.5$, moderate $\theta=3$, strong $\theta=100)$,
\item heterogenous dependency between pair of Gaussian variables (strong-weak - $\rho=(0.8,0.1)$ and weak-weak - $\rho=(0.2,0.1)$).
\end{itemize}

\begin{figure}[H]
\centering
\includegraphics[scale=0.1]{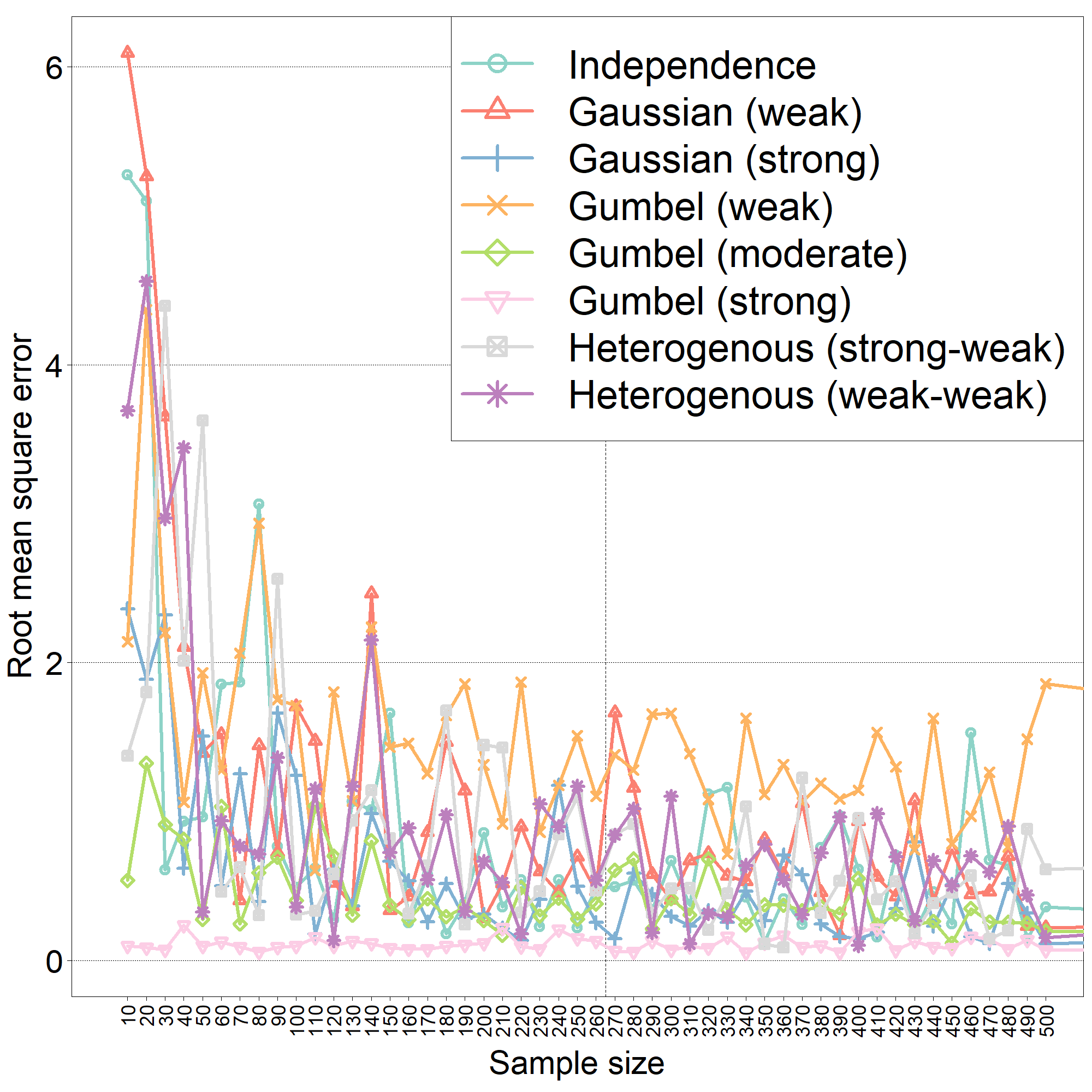}
\caption{Root mean square error of the probability-score for different dependency structure according to the sample size. The vertical dotted line indicates a sample size equals to the considered dataset ($265$).}
\label{fig:errrorDependencyStructure}
\end{figure}

Figure~\ref{fig:errrorDependencyStructure} is used to test if the estimation of the probability-score is generally accurate. For each dependency structure, samples of different size are generated. Then for each sample, the root mean square error (see Equation~\eqref{eq:rmse}) is computed using the dependency structure to compute the "true probability-score" and a fitted vine copula as the "estimated probability-score". For a sample size greater than $150$, the root mean square error is lesser than $2\%$: the probability-score remains valid for these dependence.

The White-test, with a $p$-value of $0.811$, supports the assumption that the vine-copula is a good estimator of the joint distribution of the data. Bootstrap experiments reveal that the maximum probability-score deviation falls below $6\%$ once more than $100$ colonies are used, indicating that our approach can be reliably applied to datasets of comparable size. Several apiaries contain fewer than 20 colonies after discarding dead or missing records, which reduces the reliability of the copula parameter estimates. Moreover, the number of bivariate copulas that can be reliably estimated is linked to the size of the dataset. Thus, future studies should aim to increase the number of monitored colonies per apiary or monitored apiaries, to improve the reliability and the adaptability of the model.

For the present Melior'Api dataset, accounting for dependence changes the probabilistic calibration only moderately, because dependence between the selected traits is weak. The framework nevertheless provides a general mechanism that remains valid when stronger or nonlinear dependence occurs.

The current ranking is based solely on phenotypic measurements. Incorporating genotypic information (e.g., marker-based breeding values) into the multivariate model could further enhance selection accuracy, an avenue that aligns with the goals of the Melior'Api project.

\subsection{Probability-score computation}

In the following, the results with the fitted copula are compared to the results when no information are known on the data's distribution. 

As explained in Section~\ref{subsubsec:risk}, the uniform distribution (\emph{i.e.} independence copula) is used to model the data when no information are known. The ranking criterion based on the weighted distance (see~\ref{subsubsec:weightD} is transformed into a probability-score by considering the volume of a $K$-dimensional ellipsoid (Equation~\ref{eq:riskIndependant}) for both the data standardised according to the min-max scaling (Equation~\ref{eq:minMaxScaling}) and the rank scaling (Equation~\ref{eq:rankScaling}). The former allows a comparison to a natural standardisation of the data, the latter allows a verification of the interest of the fitted copula. In both cases, each apiary is considered a unique dataset

For the fitted copula, the apiaries are fused in a complete dataset following Section~\ref{subsubsec:tutorial}. The vine copula, fitted using the function "vinecop" of the library "rvinecopulib" on R version 4.5.2, is described in Table~\ref{tab:fittedCopula}. The method is designed to accommodate dependence, but in the present dataset the estimated dependence is weak, so the probabilistic calibration differs relatively little from the independence model.

\begin{table}[t]
\centering
  \caption{Description of the fitted vine copula on the complete dataset (1 - Harvest, 2 - Strength, 3 - Hygiene, 4 - Infestation).}
{\scriptsize
\begin{tabular}{c*{11}{c}}
\hline
tree & edge & conditioned & conditioning & var\_types & family & rotation & parameters & df & tau & loglik\\
\hline
1 & 1 & 1, 2 &  & c,c & indep & 0 &  & 0 & 0.000 & 0.000\\
1 & 2 & 3, 2 &  & c,c & joe & 0 & 1.122 & 1 & 0.066 & 2.382\\
1 & 3 & 2, 4 &  & c,c & clayton & 0 & 0.177 & 1 & 0.081 & 2.421\\
2 & 1 & 1, 4 & 2 & c,c & indep & 0 &  & 0 & 0.000 & 0.000\\
2 & 2 & 3, 4 & 2 & c,c & indep & 0 &  & 0 & 0.000 & 0.000\\
3 & 1 & 1, 3 & 4, 2 & c,c & indep & 0 &  & 0 & 0.000 & 0.000\\
\hline
\end{tabular}
}

\label{tab:fittedCopula}
\end{table}

In the following because the apiaries have different number of colonies, the rank of a colony is divided by the number of colonies in its apiary. It is called relative rank. Thus, for each apiary the relative rank of the colonies is between $0$, the best colony, and $1$, the worst colony, allowing an easier comparison between apiaries. 

\begin{figure}[H]
\centering
\includegraphics[scale=0.2]{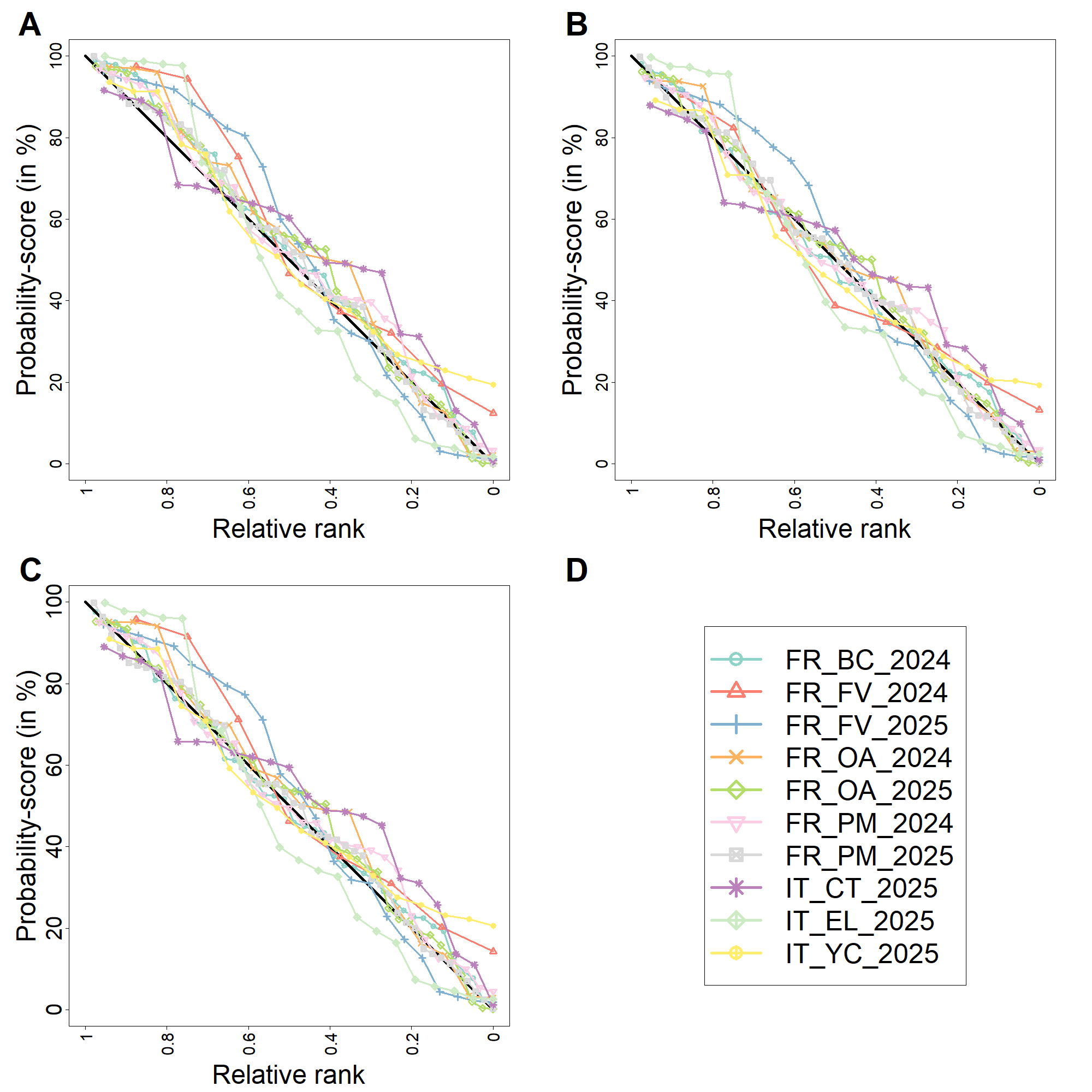}
\caption{Probability-score computed from the min-max scaling (A), the rank scaling (B) and computed with the fitted copula (C). Legend is given in (D).}
\label{fig:allRisk}
\end{figure}

Figure~\ref{fig:allRisk} displays for each apiary, the probability-score computed with the the min-max scaling (A), the rank scaling (B) and the fitted copula (C). In each case, the coloured lines are the apiaries and the black line is the evolution of the probability-score if the probability-score is obtained from the independence copula and a uniformly distributed distance. The bests colonies are located at the bottom right: lowest probability-score and lowest relative rank. Across all figures, the probability-score of an apiary is relatively similar. This figure only represents the uniform strategy (Table~\ref{tab:strategy}), the results of the other two strategies are similar.

\begin{figure}[H]
\centering
\includegraphics[scale=0.2]{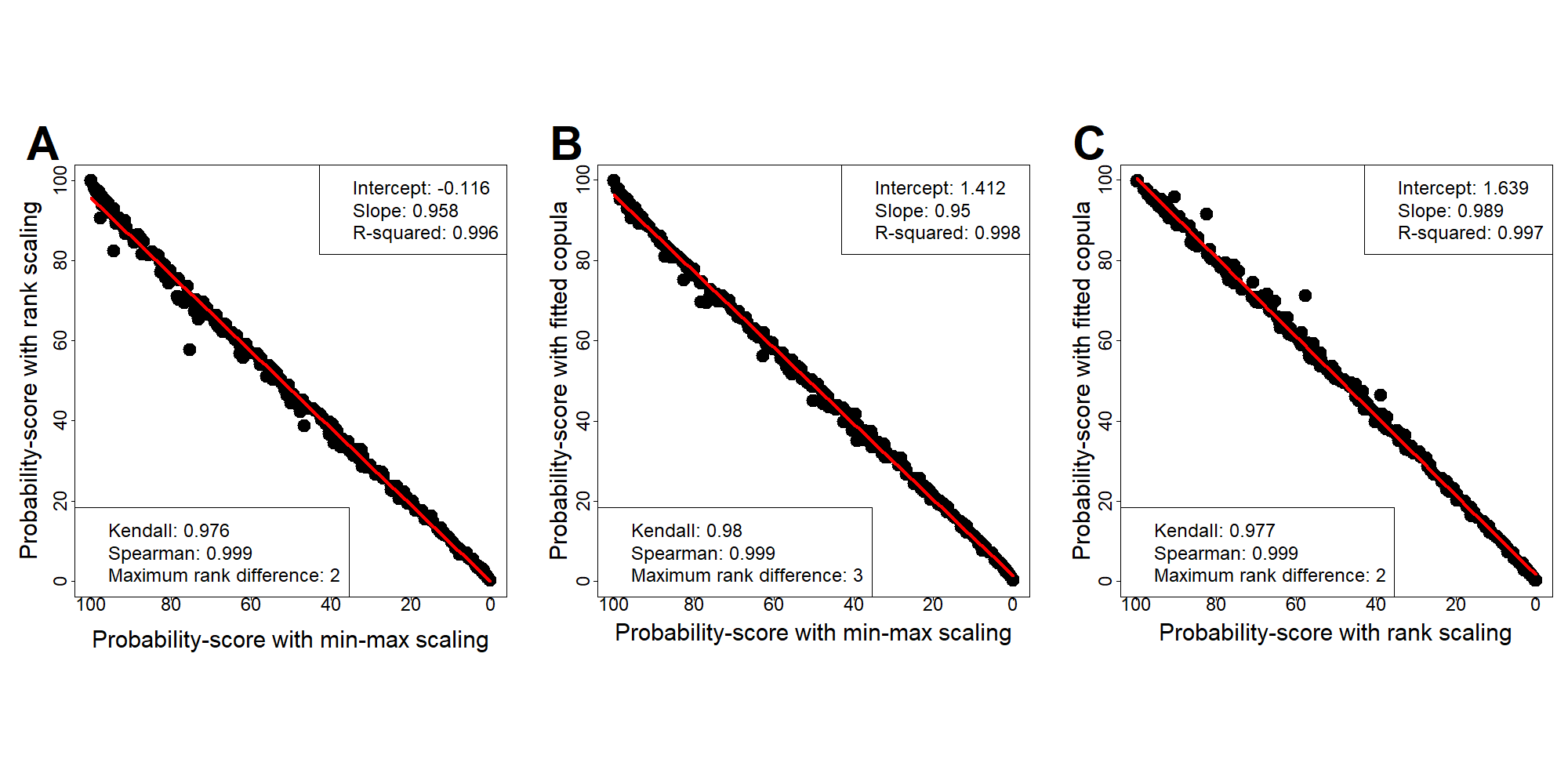}
\caption{Probability-score obtained with min-max scaling versus rank scaling (A), probability-score obtained with min-max scaling versus fitted copula (B) and probability-score obtained with rank scaling versus fitted copula (C).
  The red line corresponds to the least squares regression. The correlation between the probability-scores and the maximum difference in ranking in a apiary is indicated.}
\label{fig:linearRegressionComparison}
\end{figure}

These three methods of probability-score computations are compared using scatter plots (see Figure~\ref{fig:linearRegressionComparison}). A linear regression is applied on the probability-score obtained with min-max scaling versus rank scaling (A), the probability-score obtained with min-max scaling versus fitted copula (B) and the probability-score obtained with rank scaling versus fitted copula (C). Clearly, there is very strong positive association between the two probability scores in each case, as confirmed by tests for significant association.

%In the three cases, according to both the significativity-test and the F-test, there is significant positive relationship between the pairs of probability-scores. Moreover, in each case both the Kendall's $\tau$ and Spearman's $\rho$ correlations are close to $1$.

The probability-score relies on the joint distribution of the data and accounts for the dependence structure between variables. In the pooled dataset, only the correlation between the decrease of \emph{Varroa} mites between before the honey production of reference and the beginning of autumn ($Infestation$) and the number of bees after winter ($Strength$) is significant. This results in most sub-copulas of the fitted vine copulas being chosen as the independent copula. The probability-score resulting from both standardisation (see Equation~\eqref{eq:rankScaling} and Equation~\eqref{eq:minMaxScaling}) and the probability-score with the fitted copula are almost the same with a linear relationship with a slope of at least $0.95$ and $R^2$ of at least $0.996$.

The main advantage of the distance ranking method we propose, is the ease to define strategies, by changing the weights. This method can be generalised to any number of criteria. By giving a mathematical framework to the empirical ranking, a naive user can use the expertise of an experienced user. However, the value of the distance may be difficult to understand and interpret for people without mathematical background. The probability-score with a fitted copula does not alter the ranking of colonies based on the distance. However, since this metric is expressed as a probability of being less optimal than the perfect colony, it facilitates decision-making for non-statistical audiences by giving the percentage of being outperformed. For example, even if the value of the distance could appear relatively large, the corresponding probability-score could still be very low, especially in situations where colonies cannot be top-performing in all $K$ considered criteria.

%Three weighting strategies were evaluated (uniform, productivity biased, \emph{Varroa} biased). Changing weights reshapes the ranking, allowing beekeepers to prioritise specific behaviour. This flexibility is particularly valuable for breeding programs that must adapt to changing market demands or disease pressures. 

\subsection{Effect of the strategies}

\begin{figure}[t]
\centering
\includegraphics[scale=0.1]{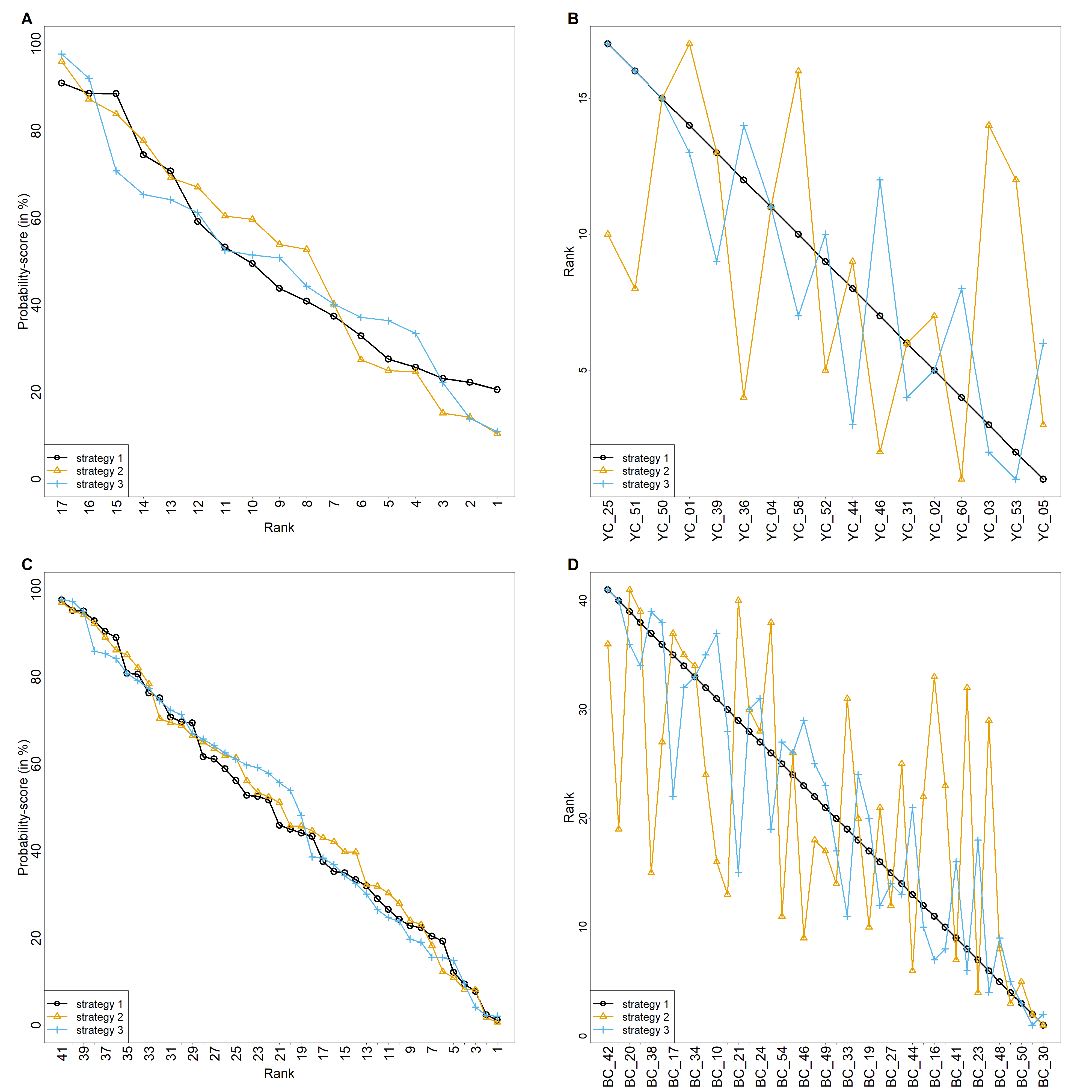}
\caption{Probability-score for apiaries IT$\_$YC$\_$2025 (A,B) and FR$\_$BC$\_$2024 (C,D). (A,C) are probability-score versus rank and (B,D) are rank for each colony.
  Each colour/line corresponds to one of the three weighting strategies described in the methods (1 - \textbf{uniform}, 2 - \textbf{productivity-focused}, and 3 - \textbf{\emph{Varroa}-focused}).}
\label{fig:riskApiary1}
\end{figure}

Figure~\ref{fig:riskApiary1} shows the probability-score according to the rank for the three strategies (A,C) and the rank of each colony for the three strategies (B,D). Only the apiaries IT$\_$YC$\_$2025 (A,B) and FR$\_$BC$\_$2024 (C,D) are represented. Even if the probability-score of each rank is close across the three strategies, the ranks of a given colony for the strategies may differ greatly.

\begin{figure}[H]
\centering
\includegraphics[scale=0.2]{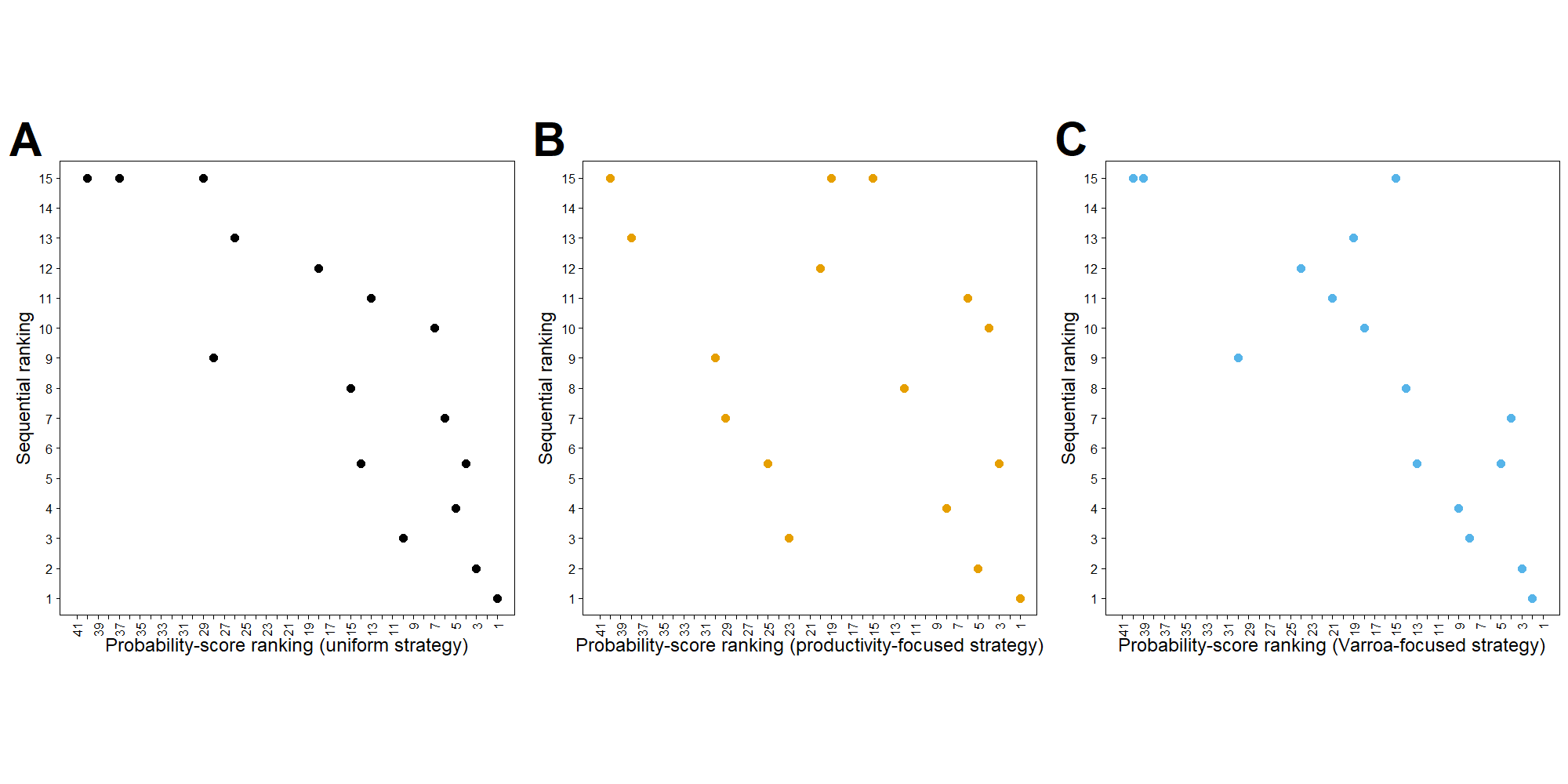}
  \caption{Comparison between the sequential ranking and the probability-score based ranking with the \textbf{uniform} strategy (A), \textbf{productivity-focused} strategy (B) and \textbf{\emph{Varroa}-focused} strategy (C).}
\label{fig:comparisonSpecialist}
\end{figure}

Figure~\ref{fig:comparisonSpecialist} compares the sequential ranking with each of the rankings resulting from the probability-score computed from the fitted copula, for apiary FR$\_$BC$\_$2024 and with the three proposed strategies. The rankings from the probability-score does not necessarily favour specialist colonies.

Table~\ref{tab:riskApiary1} gives the probability-score in percentage for each chosen strategies of apiaries IT$\_$YC$\_$2025 and FR$\_$BC$\_$2024. Only the colonies in the top $3$ of at least one chosen strategy are shown (bold values of probability-score). By construction, if the probability-score of the top ranking colonies are low, then the number of colonies in the sequential ranking is low: the best values for each criterion are spread among a low number of colonies. Reciprocally, if the best values for each criterion are spread among a high number of colonies, the sequential ranking will have a high number of individuals and the probability-score of the best ranking colonies will also be high. For each colony, the ratio of ranking where the colony is in the top $3$ over $156903$ possibles strategies is given by $P_{W,3}(i)$. For example, the colony BC$\_$15 of apiary FR$\_$BC$\_$2024 is in the top $3$ in $85.7\%$ of the possible strategies ($P_{W,3}=85.7\%$). Thus, the colony BC$\_$15 is a robust generalists. For colony YC$\_$44 $P_{W,3}=21.4\%$, this is a strategy-dependent specialist ($P_{W,3}=21.4\%$). 

\begin{table}[h]
\centering
  \caption{Probability-score (in $\%$) of the top $3$ ranked colonies of at least one of the chosen strategies for apiaries IT$\_$YC$\_$2025 and FR$\_$BC$\_$2024. The probability of being in the top $3$ over all possibles strategies is $P_{W,3}$.}
{\scriptsize
\begin{tabular}{c||c|*{3}{c}|c}
\hline
& Colony & Uniform & Productivity-focused & \emph{Varroa}-focused & $P_{W,3}(i)$\\
\hline
\multirow{6}{6em}{IT$\_$YC$\_$2025} & YC$\_$03 &\textbf{23.170} &77.780 &\textbf{14.013} & 0.502\\
&YC$\_$05 & \textbf{20.606} & \textbf{15.178} & 37.220 & 0.483\\
&YC$\_$44 & 40.908 & 53.949 & \textbf{22.118} & 0.214\\
&YC$\_$46 & 37.466 & \textbf{14.246} & 61.183 & 0.225\\
&YC$\_$53 & \textbf{22.239} & 67.093 & \textbf{10.895} & 0.431\\
&YC$\_$60 & 25.705 & \textbf{10.501} & 44.349 & 0.414\\

\hline
\multirow{4}{6em}{FR$\_$BC$\_$2024} & BC$\_$15 & \textbf{2.388} & \textbf{1.709} & \textbf{2.099} & 0.857\\
& BC$\_$22 & 9.506 & \textbf{8.135} & 14.858 & 0.269\\
& BC$\_$30 & \textbf{1.186} & \textbf{0.727}  & \textbf{2.296} & 0.890\\
& BC$\_$50 & \textbf{7.778} & 10.964  & \textbf{4.171} & 0.397\\
\hline
\end{tabular}
}

\label{tab:riskApiary1}
\end{table}

%The probability-score relies on the joint distribution of the data and accounts for the dependence structure between variables. Even if the ranking is mainly the result of the distance and the choice of the weights, the proposed score introduces biological and statistical information and interpretability in the ranking. 

\subsection{Bootstrap study of the probability-score}

Bootstrap samples~\citep{LiTanZhoPho15} are used to study the robustness of the probability-score. 

The goodness-of-fit of the fitted vine copula is assessed using the test based on White's information matrix equality proposed in~\citep{Sch19} and implemented  in the R library "VineCopula". By using $10^4$ bootstrap steps, the test statistics is $0.0336$ with a bootstrap $p$-value of $0.811$. Thus, the null hypothesis of correct model specification is not rejected at the $5\%$ significance level, suggesting that the fitted vine copula provides an adequate representation of the dependence structure in the data.

\begin{figure}[H]
\centering
\includegraphics[scale=0.2]{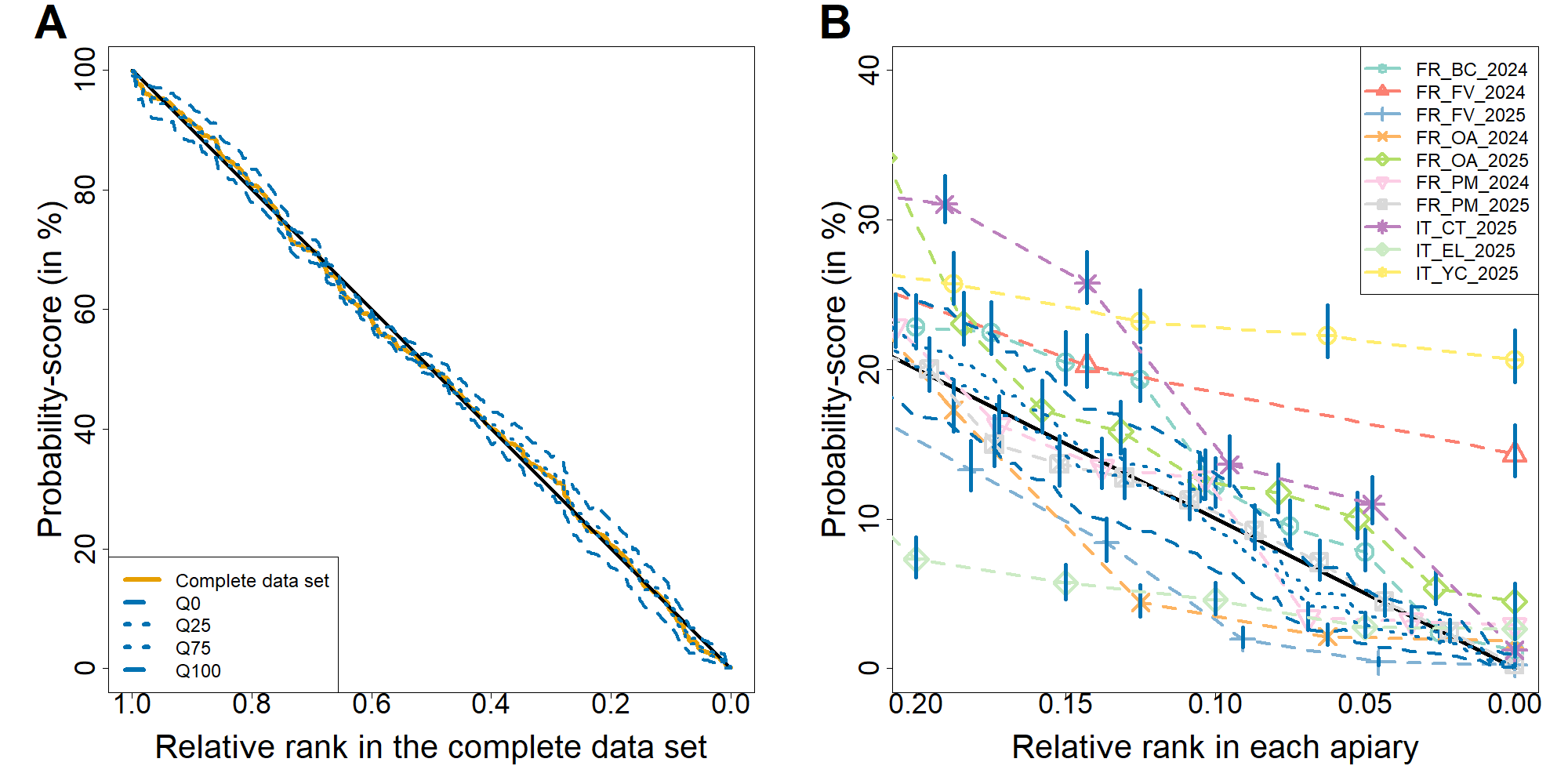}
\caption{Quantile of the probability-score computed on $10^4$ bootstrap samples of size $260$ according to the relative rank (A) and $5\%$-quantile and $95\%$-quantile of the probability-score computed from the bootstrap samples for the top-ranked colonies of each apiary (B).}
\label{fig:riskCompletBootstrap}
\end{figure}

Figure~\ref{fig:riskCompletBootstrap} shows the probability-score computed on the complete dataset and an envelope made by the probability-score computed on $10^4$ bootstrap samples with replacement of size $260$, a size similar to the complete dataset ($265$). More specifically, the envelope is made colony by colony. For a given colony, this envelope takes into consideration all the probability-score computed on the bootstrap samples containing it. On (A), the probability-score of each colony obtained with the complete dataset is represented in orange, the dotted blue lines are the quantiles computed on the bootstrap samples. The highest absolute difference between the probability-score computed on the complete data of a colony and one computed on a bootstrap sample is $4.249\%$. On (B), only the $20\%$ of top-ranked colonies of each apiary are represented. For each colony, the $5\%$-quantile and the $95\%$-quantile of the probability-score computed from the bootstrap samples are represented in blue. The probability-score is relatively stable for a fixed sample size.

\begin{figure}[H]
\centering
\includegraphics[scale=0.1]{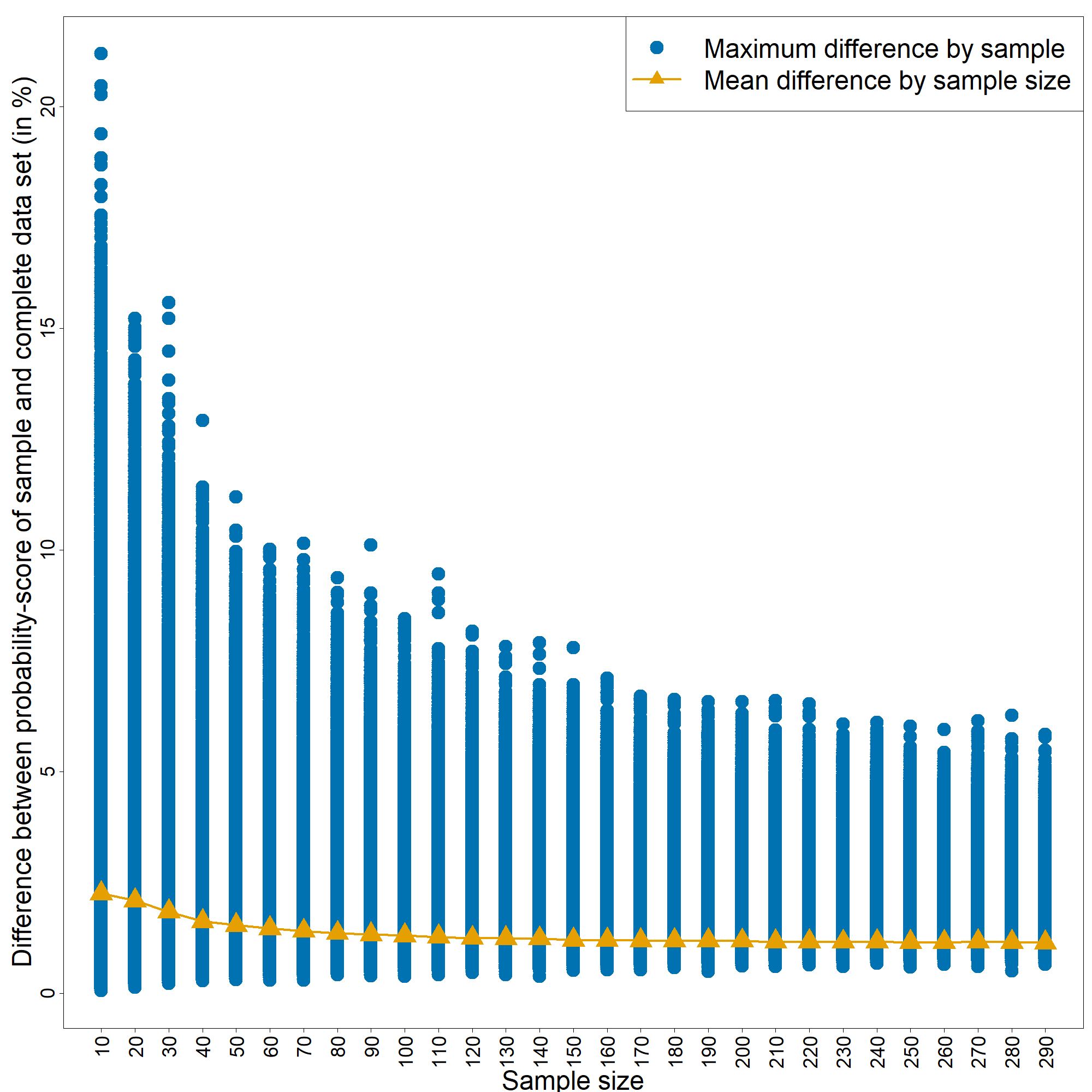}
\caption{Absolute difference in probability-score between a bootstrap sample and the complete dataset. Blue dots are the maximum difference in probability-score in a bootstrap sample. Orange triangles are the mean difference between the samples of a set size and the complete dataset.}
\label{fig:riskBootstrap}
\end{figure}

Figure~\ref{fig:riskBootstrap} shows the previous bootstrap sampling applied with different sample size. The maximum difference between the probability-score computed on the complete dataset and the probability-score computed on a bootstrap sample is represented by a blue dot. The average difference between the probability-score computed on the complete dataset and the probability-score computed on bootstrap sample of a given size is represented by an orange triangle. The probability-score estimate starts stabilising at size $100$ with a maximum difference in probability-score of less than $6\%$ and an average difference in probability-score of around $1\%$. When sample size approaches the full dataset ($265$) at size $200$, the maximum difference in probability-score is lower than $4\%$ and the average difference in probability-score is around $0.5\%$. This suggests that the method can be reliably applied to datasets of at least $100$ colonies.

\begin{figure}[H]
\centering
\includegraphics[scale=0.20]{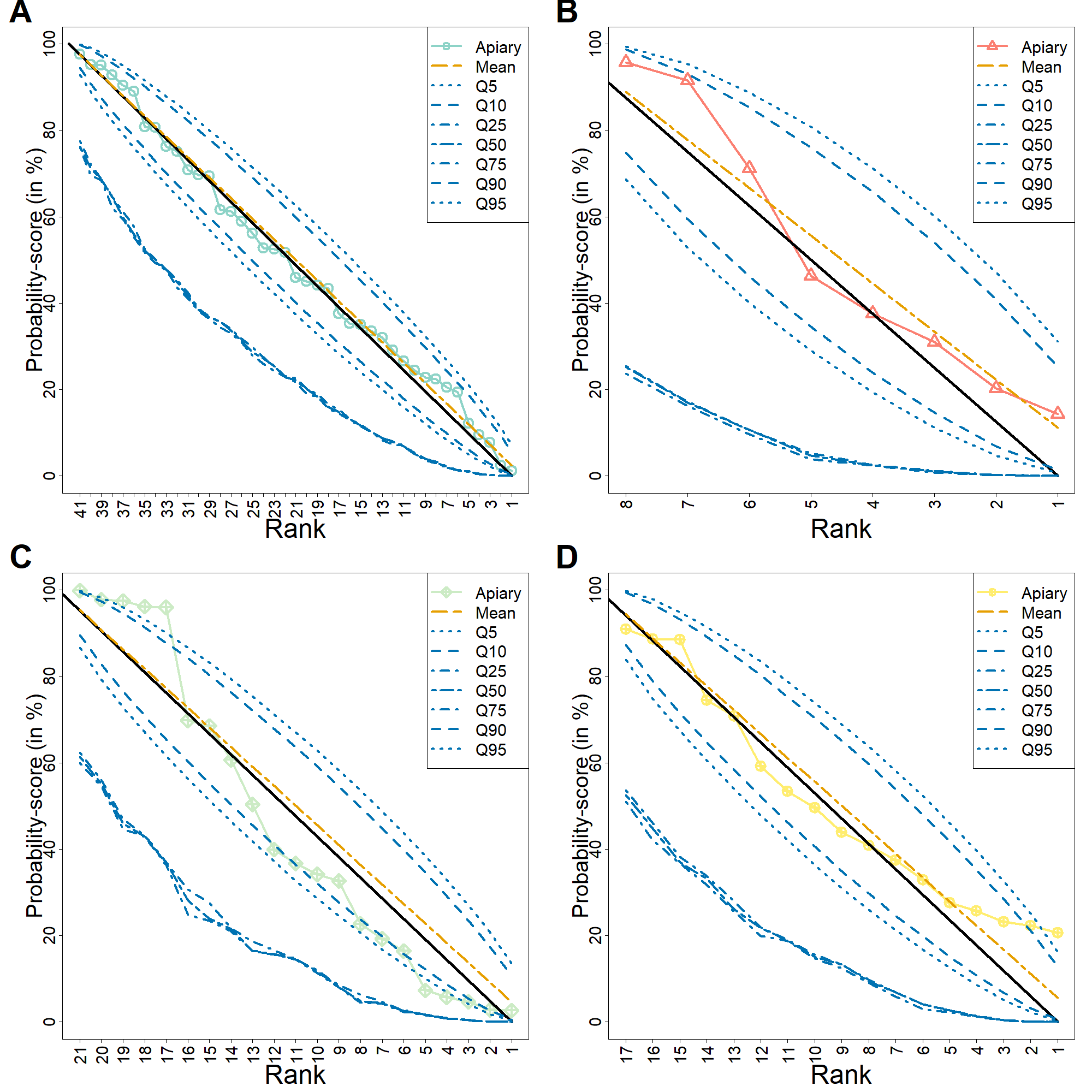}
\caption{Envelope test for four apiaries (FR$\_$BC$\_$2024 - A, FR$\_$FV$\_$2024 - B, IT$\_$EL$\_$2025 - C, IT$\_$YC$\_$2025 - D). The quantile obtained from $10^4$ the simulated apiaries are in blue. The black line is the probability-score curve of the observed apiary.}
\label{fig:riskRandomApiary1}
\end{figure}

As another diagnostic to check whether pooling all apiaries for the estimation of the copula is justified consists of an envelope tes, applied on each apiary to assess whether the observed data could have been generated by the fitted vine copula (see Figure~\ref{fig:riskRandomApiary1} and Appendix for the rest of the apiaries). The copula is fitted on the complete dataset. For each apiary size, $10^4$ samples of same size are generated from the fitted copula. The probability-score of these samples are used to generate an envelope (blue dotted lines). The mean value of the probability-score of these samples are represented by the orange dotted line. Except for the Italian apiaries IT$\_$EL$\_$2025 and IT$\_$YC$\_$2025, the observed probability-score curves remain within the $5\%$-$95\%$ envelope, indicating that the vine copula provides an adequate generative model for the majority of apiaries and confirming the suitability of the pair copula decomposition. For IT$\_$EL$\_$2025 the three lowest ranked colonies (ranks $17$ - $19$) have probability-score values above the $95\%$ envelope, \emph{i.e.} they are considerably worse than what would be expected under the fitted model. Conversely, for IT$\_$YC$\_$2025 the rank $1$ colony exhibits a probability-score higher than the $95\%$ envelope, suggesting that this colony has unexpectedly badly performed.

\begin{figure}[H]
\centering
\includegraphics[scale=0.1]{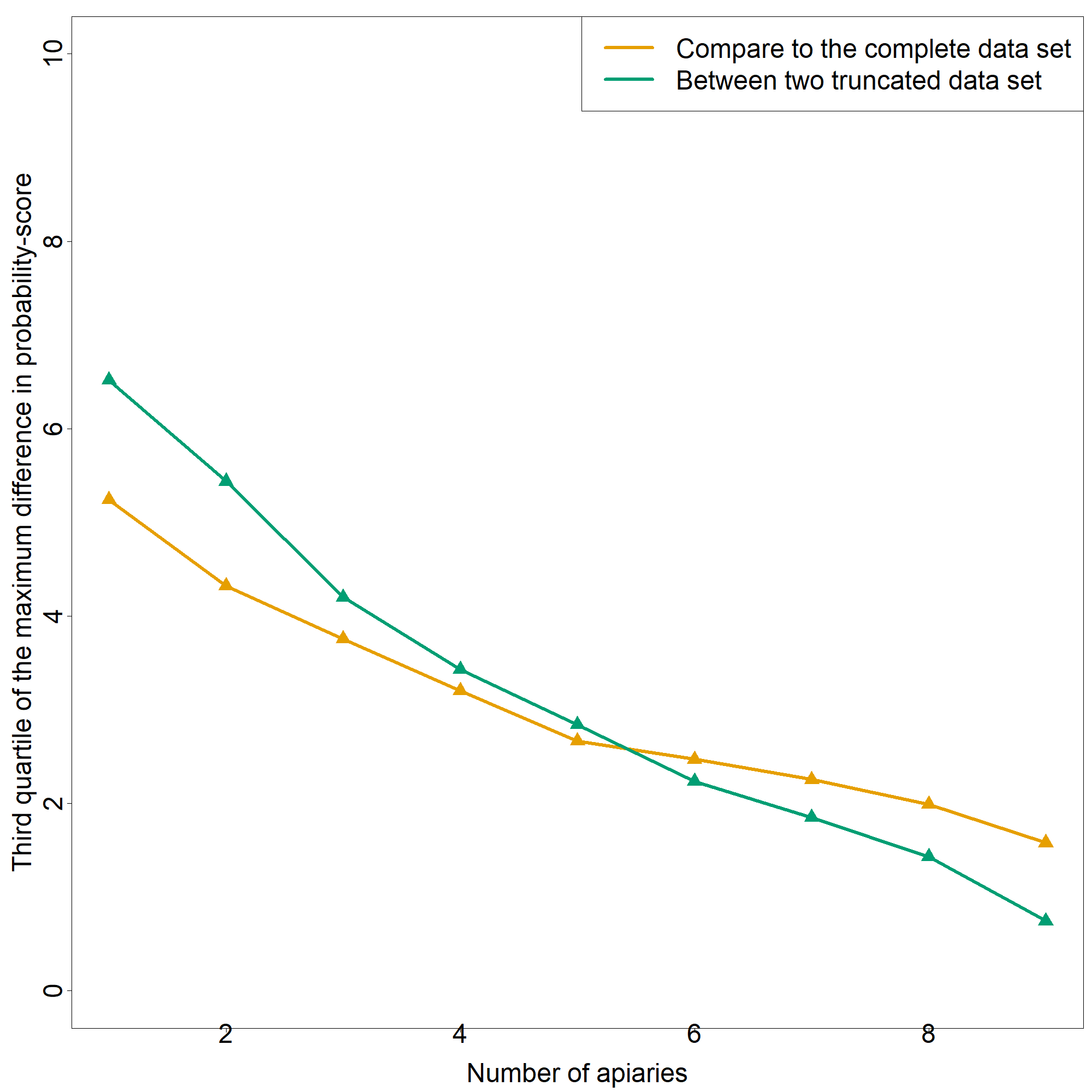}
\caption{Third quartile of the maximum difference between the probability-score computed on the complete dataset and the probability-score computed on a truncated dataset (orange) and between two truncated dataset (green) over every combination of a given number of apiaries.}
\label{fig:leaveOutApiaries}
\end{figure}

Because the apiaries are pooled, the fitted copula should model the relative performance of the colonies within apiaries. Figure~\ref{fig:leaveOutApiaries} shows the effect of the apiaries in the estimation through a leave-apiaries-out validation. For all number of apiaries, every combination is used to create truncated datasets. Then the probability-score is computated on each truncated dataset. These probability-scores are compared to the probability-score computated on the complete dataset (orange) and to every other truncated datasets (green). In the Figure and for a given number of apiaries, the third quartile of the maximum difference in probability-scores for every combination is represented. As expected, the higher the number of apiaries, the lower the difference in probability-score. Thus, the copula is able to estimate the relative performance of the colonies within an apiary.

\section{Conclusion}

\label{sec:conclusion}

This paper introduces a multi-criteria ranking framework that improves interpretability and calibrates the multi-criteria distance using the dependence structure of the reference population. This method is used for honeybee breeder colony selection.

This framework combines a breeder-defined multi-criteria decision rule based on a weighted Euclidean distance to an ideal colony with a statistical model of the joint distribution of the criteria. The distance provides the ordering of colonies, whereas the fitted joint distribution transforms this distance into a population-relative probability-score. The probability-score therefore provides an interpretable measure of the proportion of reference-population colonies expected to outperform a candidate under the selected breeding objective.

Future work will focus on integrating genotypic information (e.g. marker based breeding values) to enrich the multivariate model, and on exploring meta-ranking techniques such as score fusion to combine multiple weighting strategies~\citep{EicVriCol13}.  

A user-friendly RShiny interface is currently in development. It is available at \url{https://forge.inrae.fr/christophe.reype/Inrae_Bees}.

\bibliography{reference}% common bib file

\begin{appendices}

\label{sec:appendix}

\begin{figure}[t]
\centering
\includegraphics[scale=0.2]{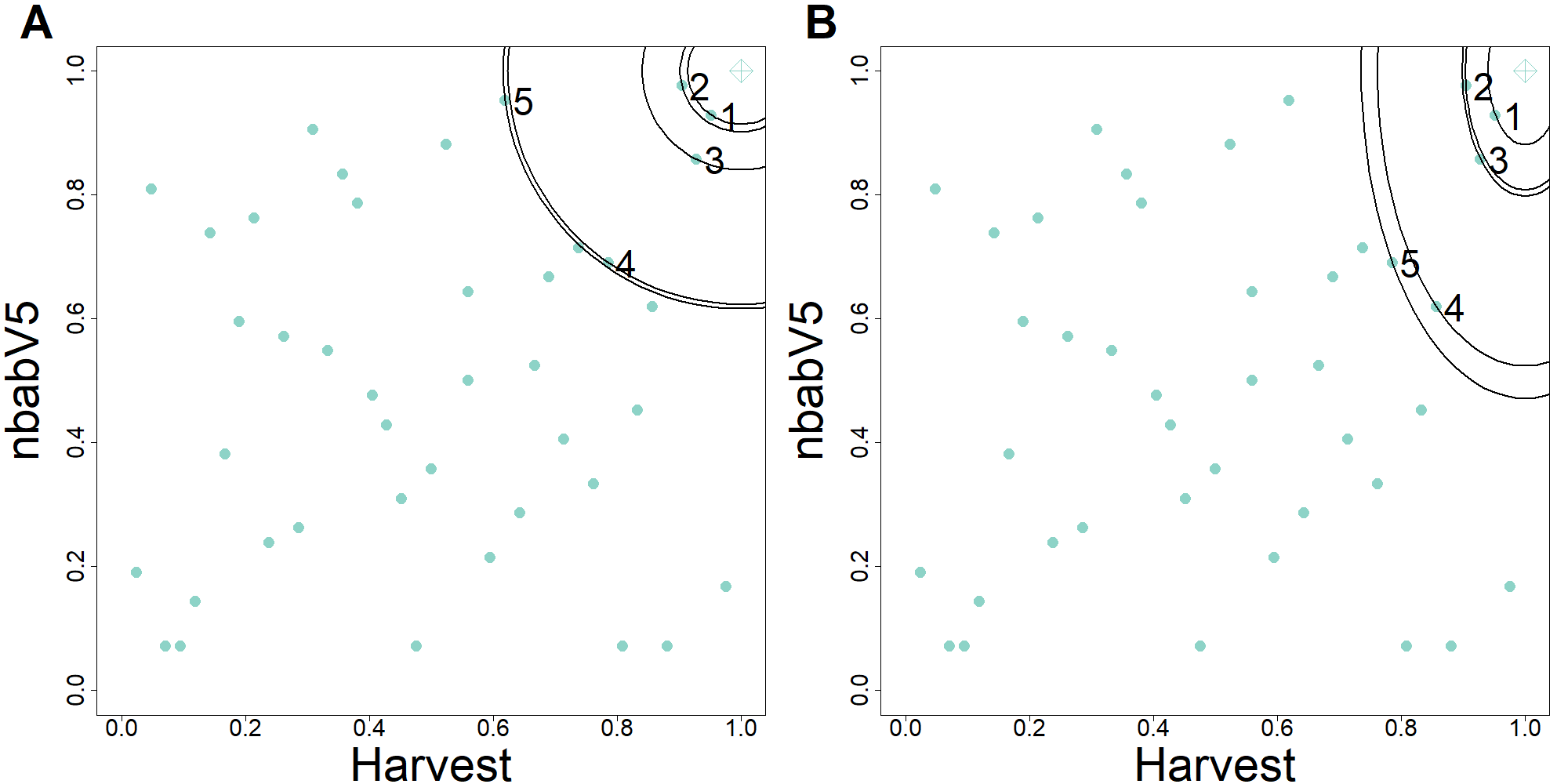}
\caption{Effect of the strategy ($w=(0.5,0.5)$ - A, $w=(0.8,0.2)$ - B) on the projection of the apiary FR$\_$BC$\_$2024 in the bi-dimensional space formed by $Harvest$ and $Strength$ with the min-max scaling. The rank and the equidistant points of the top $5$ colonies are indicated.}
\label{fig:effectPond2}
\end{figure}

\begin{figure}[t]
\centering
\includegraphics[scale=0.20]{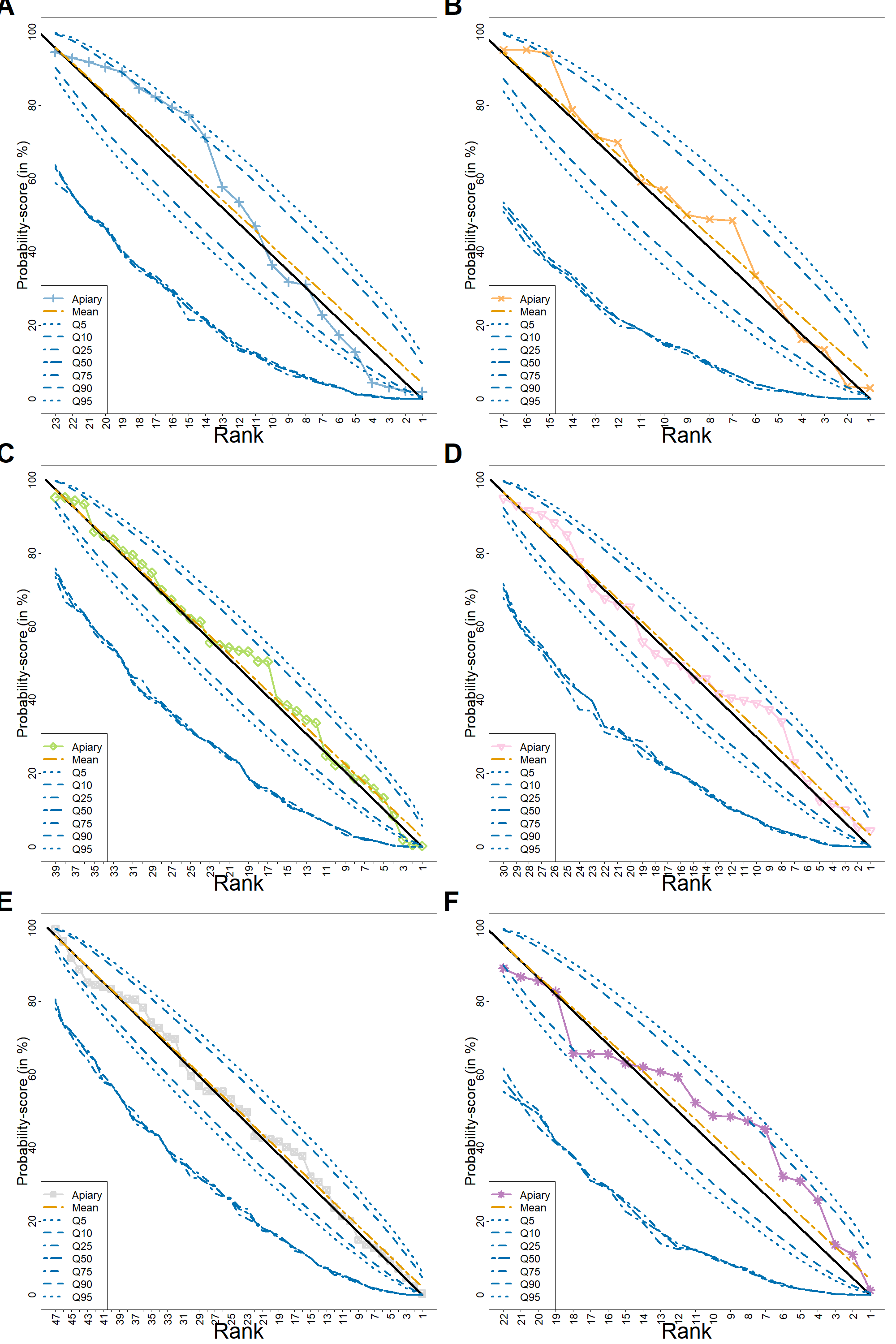}
\caption{Envelope test for the other apiaries (FR$\_$FV$\_$2025 - A, FR$\_$OA$\_$2024 - B, FR$\_$OA$\_$2025 - C, FR$\_$PM$\_$2024 - D, FR$\_$PM$\_$2025 - E, IT$\_$CT$\_$2025 - F).  The quantile obtained from $10^4$ the simulated apiaries are in blue. The black line is the probability-score curve of the observed apiary.}
\label{fig:riskRandomApiary2}
\end{figure}

\end{appendices}

%% file, and delete the associated \verb+\bibliography+ commands. &  &     %%
%%===========================================================================================%%

\end{document}